\documentclass[a4paper,french]{rnti}

\usepackage[T1]{fontenc}
\usepackage[utf8]{inputenc}
\usepackage[euler]{textgreek}
\usepackage{caption}
\usepackage{url}
\usepackage{graphicx}
\usepackage{xcolor}

\usepackage{tabularx, booktabs}

\usepackage{makecell}
\usepackage{multirow}

\titrecourt{Évaluation de LLM pour les historiens}

\titre{Évaluation des capacités de réponse de larges modèles de langage (LLM) pour des questions d'historiens}

\nomcourt{M. Chartier et al.}

\auteur{
        Mathieu Chartier\affil{1},
        Nabil Dakkoune\affil{2},\\
        Guillaume Bourgeois\affil{1} et
        Stéphane Jean\affil{2}
        }

\affiliation{
    \affil{1}CRIHAM, université de Poitiers\\
    \affil{2}LIAS, ISAE-ENSMA et université de Poitiers\\
        prenom.nom@univ-poitiers.fr\\
 }
 
\resume{Les larges modèles de langage (LLM) tels que ChatGPT ou Bard ont bouleversé la recherche d’informations et conquis le public par leur facilité à générer des réponses sur mesure en un temps record, qu’importe le sujet. Dans cet article, nous analysons les capacités de différents LLM à produire des réponses sur des faits historiques en français avec fiabilité, exhaustivité et suffisamment de pertinence pour être directement exploitables ou extractibles. Pour cela, nous avons élaboré un banc d’essai constitué de multiples questions d’histoire. Ces dernières sont de différents types, thématiques et de niveaux de difficulté variables. Notre évaluation des réponses fournies par dix LLM, que nous avons jugés pertinents, montre de nombreuses limites sur le fond comme dans la forme. Au-delà d'un taux de précision globalement insuffisant, nous mettons en évidence le traitement inégal du français ou encore des problèmes de loquacité et d’inconstance des réponses fournies par les LLM.}

\summary {Large Language Models (LLMs) like ChatGPT or Bard have revolutionized information retrieval and captivated the audience with their ability to generate custom responses in record time, regardless of the topic. In this article, we assess the capabilities of various LLMs in producing reliable, comprehensive, and sufficiently relevant responses about historical facts in French. To achieve this, we constructed a testbed comprising numerous history-related questions of varying types, themes, and levels of difficulty. Our evaluation of responses from ten selected LLMs reveals numerous shortcomings in both substance and form. Beyond an overall insufficient accuracy rate, we highlight uneven treatment of the French language, as well as issues related to verbosity and inconsistency in the responses provided by LLMs.}

\begin{document}
\bibliographystyle{rnti}

\section{Introduction}

Le succès et l’engouement rencontrés par ChatGPT d’OpenAI depuis son lancement le 30 novembre 2022 mettent en évidence les progrès des larges modèles de langage (LLM) auprès du grand public, atteignant un million d’utilisateurs actifs seulement cinq jours après l’annonce officielle et même cent millions deux mois plus tard. Les LLM démontrent des capacités en termes de production et complétion de textes, de traduction automatique mais aussi de génération de réponses à des questions de tout ordre. Ces outils se révèlent notamment performants dans ces différentes tâches grâce à un entraînement non supervisé ou semi-supervisé sur d’immenses corpus textuels, totalisant des milliards de mots parmi des sources diverses et~variées. 

Avec de telles masses d’informations ingurgitées, nous pouvons estimer que ces modèles sont entraînés sur des volumes de connaissances paraissant dépasser les capacités mémorielles humaines (\cite{bowman_eight_2023}), laissant penser à tort qu’ils auraient réponse à tout. En effet, bien que les LLM s’appuient sur des millions de sources diversifiées, ils sont capables de produire avec une certaine assurance des réponses biaisées voire totalement fausses ou farfelues (\cite{zheng_why_2023}), mais aussi développer un certain niveau d’hallucination (\cite{ji_survey_2023}). Par conséquent, il semble nécessaire d’analyser leur aptitude à générer des réponses approfondies et exactes, validées par des spécialistes, notamment dans des domaines aussi complexes que vastes tels que l’histoire ou plus largement les sciences humaines et sociales qui font appel à des connaissances diachroniques de plus en plus fouillées et nombreuses.

Les spécialistes des sciences humaines et sociales, et plus spécifiquement les historiens, ont pour habitude d’effectuer des recherches approfondies au sein de vastes ensembles documentaires ou archivistiques. Pour les accompagner et les aider à trouver plus facilement des sources adaptées à leurs travaux, de nombreux outils numériques ont vu le jour tels que Gallica\footnote{\http{https://gallica.bnf.fr/}} ou Europeana\footnote{\http{https://www.europeana.eu/fr}}. Cependant, ces ressources numériques présentent deux limites majeures pour les spécialistes : d’une part, les données ne sont pas toujours disponibles en ligne, soit parce que les documents ne sont accessibles que physiquement, soit parce que leur numérisation n’est que partielle ; d’autre part, ces outils ne recouvrent pas la complexité des problématiques historiques et proposent uniquement des listes de ressources résultant de requêtes de mots clés souvent ambiguës voire mal traitées informatiquement. 

À titre d’exemple, pour la requête courte « roi de France », certains outils cherchent des ressources qui contiennent soit « roi », soit « France », ou pire, le mot vide « de », engendrant ainsi une liste interminable de résultats potentiellement imprécis et inappropriés. Même pour des requêtes simples comme « date du siège de La Rochelle », un site web de qualité comme Gallica va retourner des documents pertinents comme \textit{le Siège de La Rochelle : journal contemporain (20 juillet 1627-4 juillet 1630)}, mais aussi de nombreux documents inappropriés comme \textit{Recueil de pièces en prose et en vers, lues dans les assemblées publiques de l'Académie royale des belles-lettres de La Rochelle} ou \textit{L'aire urbaine de La Rochelle plus dynamique que ses cons\oe urs du littoral}. Par conséquent, là où les historiens aimeraient facilement obtenir des amorces de réponses précises et sourcées à l'aune de leurs propres orientations, la majorité des outils existants s'en tiennent à des ressources cataloguées qu’il convient de consulter et d’analyser en détail par d’autres moyens. Il nous semble donc intéressant de déterminer si les LLM disposent de données et de facultés suffisantes pour aider les historiens dans leurs recherches.

Dans cet article, nous analysons les capacités des LLM dans la génération de réponses pertinentes et complètes pour différents types de requête en histoire. Nous considérons ainsi des requêtes quantitatives (par exemple : « Combien de batailles ont eu lieu au cours de la troisième guerre de religion ? ») et qualitatives (cela peut être une demande fermée comme « Quelle est la date de création de la forge de Verrières à Lhommaizé ? » ou une question ouverte telle que « Dans quelle mesure l'artisanat a été essentiel dans la vie économique des villages du Poitou de 1500 à 1800 ? »). Nous testons également différentes complexités de requêtes allant de questions rationnelles et simples à de profondes interrogations d’historien à la fois composites et équivoques. En effet, il est certainement plus simple de fournir la date et le lieu de naissance exacts d'un illustre personnage historique que d'apporter une réponse via un raisonnement dialectique sur un sujet multiforme et ambigu du passé. Nos travaux évaluent également les possibilités des LLM à produire des réponses rigoureuses pour des faits historiques aux caractéristiques variables, notamment lorsque les sujets évoqués sont peu traités en ligne ou méconnus, même de la part d’historiens spécialistes. 

Pour réaliser notre évaluation, nous proposons un banc d'essai composé de 62 questions. Pour chacune de ces questions, nous avons caractérisé les types de réponse possibles (bonne réponse, réponse partielle, etc.). Nous avons ensuite sélectionné dix LLM qui nous ont semblé être les plus pertinents pour notre étude. L'évaluation des réponses fournies par ces LLM à notre banc d'essai montre de nombreuses limites sur le fond comme dans la forme. Au-delà d'un taux de précision globalement insuffisant, cette évaluation met en évidence des lacunes actuelles des LLM eu égard à notre sujet tels que le traitement inégal du français ou l'inconstance des réponses fournies par les LLM.

Cet article est organisé comme suit. Dans la section~\ref{section-travaux-connexes}, nous analysons les autres travaux s'étant intéressés aux LLM et plus particulièrement sur le domaine de l'histoire. Dans la section~\ref{section-requete} nous présentons notre banc d'essai en justifiant et illustrant les questions choisies. La section~\ref{section-methodologie} présente la méthodologie suivie pour évaluer les LLM sélectionnés. Les résultats de cette évaluation sont présentés dans la section~\ref{section-resultat}. Nous concluons dans la section~\ref{section-conclusion} et introduisons des perspectives à nos travaux.

\section{Travaux connexes}
\label{section-travaux-connexes}

Plusieurs évaluations et comparatifs de performances de LLM ont déjà été réalisés dans des domaines comme les mathématiques (\cite{frieder_mathematical_2023, yuan_how_2023, hendrycks_measuring_2021}), la programmation informatique (\cite{chen_evaluating_2021, austin_program_2021}), la génération de résumés de textes (\cite{goyal_news_2023}) ou même dans la réussite potentielle de tests universitaires (\cite{nunes_evaluating_2023}). En outre, des scientifiques ont privilégié une approche holistique et multitâche, offrant ainsi une évaluation plus globale et précise des modèles de langages en anglais (\cite{lewkowycz_beyond_2022, suzgun_challenging_2022, liang_holistic_2022}) ou en chinois (\cite{huang_c-eval_2023}). Malheureusement, peu de travaux similaires s’intéressent spécifiquement aux humanités numériques, qui plus est en langue française.

Pour aller plus loin, des chercheurs ont évalué les LLM dans leurs capacités intrinsèques à répondre à des questions ou à suivre des raisonnements profonds (\cite{lin_truthfulqa_2022, stelmakh_asqa_2023, tan_evaluation_2023, guo_how_2023}). Une fois encore, ces études reposent uniquement sur la langue anglaise, sans s’intéresser à la thématique des questions historiques. La seule approche qui se penche sur la génération de questions-réponses dans les humanités numériques, et en langue française, est celle de \cite{bechet_question_2022}. Cependant, cette dernière ne s’interroge pas sur la capacité des LLM à répondre correctement à des questions historiques, mais plutôt à la faculté de générer des réponses de qualité à partir de l’extraction d’informations dans un corpus textuel historique. Notre étude prend donc le parti de tester les connaissances natives des LLM dans la production de réponses précises, complètes et fiables en langue française dans le domaine de l’histoire.

\section{Les requêtes et modèles testés}
\label{section-requete}
Les historiens cherchent autant à soutirer des réponses factuelles à des questions fermées que des explications circonstanciées pour de vastes interrogations. Nous devions donc couvrir l'ensemble des possibilités afin d'obtenir par exemple une date précise, une lignée familiale, une durée, un contexte historique ou encore une réponse approfondie sur une problématique complexe. Pour ce faire, nous avons formé diverses classes de requêtes fermées et ouvertes pour analyser pleinement les connaissances des LLM et examiner la qualité des réponses lorsque le niveau de difficulté et le type de questions diffèrent. Le tableau~\ref{Tableau1} illustre ces différents types de requêtes. Aussi, nous avons créé plusieurs groupes de requêtes quantitatives et qualitatives qui attendent des types de réponses différents afin d'évaluer les potentielles lacunes des LLM dans des conditions variables. En particulier, nous avons subdivisé les requêtes qualitatives en diverses sous-classes pour obtenir des réponses caractéristiques telles que des métadonnées (date d'un événement, nom d'un acteur historique, etc.) ou des listes d'informations sur des questions simples jusqu'à des descriptions de plus en plus détaillées sur des questionnements ouverts et complexes.

\setlength{\tabcolsep}{5pt}
\begin{table}[htb]
\small\centering
\begin{tabularx}{\linewidth}{>{\hsize=.6\hsize\centering}X>{\hsize=.4\hsize\centering}X>{\hsize=2\hsize}X}
\toprule[1.0pt]\toprule[1.0pt]
\textbf{\makecell{Types de question\\\textit{Données attendues}}} & \makecell{\textbf{Nombre}\\\textbf{de questions}} & \makecell{\textbf{Exemples de questions posées}} \\
\toprule[1.0pt]
\multirow{2}{*}{\makecell{Quantitatif (fermé)\\\textit{Donnée numérique}}} & \multirow{2}{*}{16} & Quelle est la durée de la bataille de Poitiers en 1356 ? \\[3pt]
 & & Quel est le nombre de pertes au cours du siège de La Rochelle ? \\
\midrule[0.1pt]
\multirow{2}{*}{\makecell{Qualitatif (fermé)\\\textit{Métadonnée}}} & \multirow{2}{*}{15} & Où est né le cardinal de Richelieu ? \\[3pt]
 & & Quand a été proclamée la première loi sur l'apprentissage ? \\
\midrule[0.1pt]
\multirow{3}{*}{\makecell{Qualitatif (fermé)\\\textit{Liste de données}}} & \multirow{3}{*}{10} & Quels sont les noms des enfants de Jean le Bon ? \\[3pt]
 & & Qui sont les principaux défenseurs de la réforme protestante lors de la troisième guerre de religion ? \\
\midrule[0.1pt]
\multirow{2}{*}{\makecell{Qualitatif (ouvert)\\\textit{Définition/Description}}} & \multirow{2}{*}{16} & Qu’est-ce que l'Edit de Nantes en 1598 ? \\[3pt]
 & & Comment les forgerons travaillaient le fer à l'époque moderne ? \\
\midrule[0.1pt]
\multirow{2}{*}{\makecell{Qualitatif (ouvert)\\\textit{Description détaillée}\\\textit{d'une problématique}}} & \multirow{3}{*}{5} & En quoi la constitution d’une Nouvelle République de La Rochelle a causé la future capitulation rochelaise ? \\[3pt]
 & & Pourquoi la mort d'Abd al-Rahman en 732 a été déterminante pour l'expansion musulmane en Europe ? \\
\bottomrule[1.0pt]
\end{tabularx}
\caption{Typologie et exemples de questions d'histoire posées aux LLM.}
\label{Tableau1}
\end{table}

Nos questionnements imposent de disposer de suffisamment de données à exploiter, à la fois en nombre mais aussi en qualité. Nous avons donc réparti nos questions parmi cinq grands sujets ou faits historiques s'étant déroulés dans l’ancien Poitou (faisant actuellement partie de la région Nouvelle-Aquitaine) à diverses époques :
\begin{itemize}
	\item Quatre faits historiques autour de guerres ou batailles décomposés ainsi : deux batailles de Poitiers (en 732 et en 1356) très traitées dans les sources, mais diversement bien connues des historiens et pouvant entraîner des confusions pour les outils de recherche ; un sujet évoquant les guerres de religion au travers de l’illustre siège de La Rochelle (en 1627-1628) plutôt bien maîtrisé par les historiens mais aussi autour de batailles un peu plus confidentielles et méconnues (Jarnac et Moncontour).
	\item Un sujet vaste au sujet de l'artisanat à l'époque moderne, globalement plus nébuleux, abstrait et complexe pour l'obtention de réponses concrètes via des outils de recherche. 
\end{itemize}

Au total, notre banc d'essai se compose de 62 questions dont chacune est énoncée à la fois sous forme de requêtes conversationnelles (en langue naturelle) et de requêtes de mots clés traditionnelles. Nous avons également décidé de proposer deux formulations pour chaque question afin de vérifier si les LLM répondent avec exactitude dans chaque cas, mais aussi pour nous assurer de leur bonne compréhension de nos demandes. Par exemple, la question fermée « Quelle est la date du siège de La Rochelle ? » a aussi été testée avec la variante « Quand s’est tenu le siège de La Rochelle ? ». Enfin, nous avons ajouté une formulation complémentaire à vingt requêtes qualitatives fermées pour tester spécifiquement la compréhension et la prise en compte de variations étymologiques ou toponymiques dans le texte. En définitive, nous avons testé 268 requêtes sur des faits historiques diversifiés (134 en langue naturelle et 134 constituées de mots clés).

Parallèlement au choix des questions, nous avons dû sélectionner les LLM à exploiter. Nous avons procédé à une large revue d’effectif afin de vérifier leurs spécificités et les besoins techniques pour les faire fonctionner correctement. Il nous a été évidemment impossible de couvrir l’ensemble des outils basés sur des modèles de langage, à la fois par limitation technique (par exemple, Claude 2 d’Anthropic\footnote{\http{https://www.anthropic.com/index/introducing-claude}} n’était disponible qu’aux Etats-Unis au moment des tests) ou linguistique (par exemple, l’outil Luminous d’Aleph Alpha ne se concentre que sur la langue allemande\footnote{\http{https://www.aleph-alpha.com/luminous}}). Toutefois, comme l’indique le tableau~\ref{Tableau2}, nous souhaitions étudier un panel suffisamment représentatif des LLM, avec des modèles basés sur GPT (\cite{openai_gpt-4_2023}), sur PaLM 2 (\cite{anil_palm_2023}), d’autres sur LLaMA (\cite{touvron_llama_2023}) ou T5 (\cite{raffel_exploring_2020}), mais aussi avec des constitutions variables de leurs corpus.

\begin{table}[htb]
\center\small
\begin{tabular*}{\linewidth}{@{\extracolsep{\fill}}cccc}
\hline\hline
\textbf{Technologie} & \textbf{Modèle} & \textbf{Nombre de paramètres} & \textbf{Type d'usage} \\
\hline
ChatGPT & GPT-4 & 170 000 milliards & Chat \\
ChatGPT & GPT-3.5-turbo & 175 milliards & Chat \\
Google Bard & PaLM 2 & ? & Chat \\
TextCortex AI & Sophos-2 & 20 milliards et plus & Chat \\
Guanaco & Guanaco-33b-GGML & 33 milliards & Chat \\
Vicuna & Vicuna-33b-v1.3 & 33 milliards & Chat \\
GPT4All & L13b-snoozy & 13 milliards & Chat \\
Koala & 13b-diff-v2 & 13 milliards & Chat \\
Vigogne & Instruct-13b & 13 milliards & Instruct \\
Falcon & Instruct-7B & 7 milliards & Instruct \\
\hline
\end{tabular*}
\caption{Liste des LLM testés sur des questions d'histoire.}
\label{Tableau2}
\end{table}

\section{Méthodologie suivie}
\label{section-methodologie}
Nous avons examiné manuellement, avec des spécialistes, les réponses fournies par les différents outils testés. Nous avons procédé de manière itérative en couvrant l’ensemble des types de questions d’histoire de notre sélection, c’est-à-dire en demandant aux LLM de répondre à des questions quantitatives, qualitatives fermées et ouvertes tantôt descriptives, tantôt ambiguës et complexes.

Certains modèles comme GPT, Bard ou Vigogne comprennent et couvrent parfaitement les requêtes en langue naturelle mais il nous a semblé pertinent de tester également les mêmes sujets sous forme de requêtes de mots clés traditionnelles afin de comparer les résultats obtenus, selon les deux approches, pour l’ensemble des outils analysés. En outre, nous avons répété ces questionnements deux fois pour chaque batterie de tests afin de s’assurer que les nouvelles réponses proposées par les LLM demeurent cohérentes ou apportent plus ou moins de détails au sujet des faits historiques sondés. Au total, nous avons donc analysé 5360 réponses résultant de nos 62 questions d’histoire afin d’obtenir un échantillon suffisamment représentatif des performances des dix LLM testés.

Pour mener à bien notre analyse et obtenir les résultats cités dans cet article, nous avons collecté la totalité des réponses pour l’ensemble des essais réalisés. Pour faciliter la reproductibilité de notre expérimentation, l'ensemble des questions et réponses obtenues sont disponibles sur \url{https://github.com/lias-laboratory/experimentalhistorianllm}.\\Bien qu’il existe des méthodes d’évaluation automatisée des capacités des LLM, elles montrent généralement très vite leurs limites, notamment lorsqu’il s’agit de questions ouvertes ou spécifiques à un domaine (\cite{wang_evaluating_2023, kamalloo_evaluating_2023}). En outre, une multitude de questions d’histoire font appel à des connaissances vastes et imbriquées issues de domaines aussi variés que spécifiques, il nous semblait par conséquent délicat d’opter pour des systèmes d’évaluation automatiques n’ayant pas nécessairement un niveau de connaissances suffisant. Pour ces raisons, nous avons procédé à une analyse humaine et minutieuse des réponses afin d’évaluer leur pertinence, leur précision et leur complétude grâce à des historiens et à des sources validées par les spécialistes. Notre objectif était de laisser les modèles générer des réponses, à l’image d’un élève répondant à un devoir d’histoire, et de vérifier si les notions attendues étaient bien présentes. Par exemple, pour une question quantitative fermée comme « Combien de batailles ont eu lieu au cours de la troisième guerre de religion ? », une réponse idéale devrait évoquer les huit faits majeurs tandis qu'une question ouverte telle que « Quel a été l’élément déclencheur de la bataille de Poitiers en 1356 ? » attendrait une réponse détaillée évoquant la chevauchée du Prince Noir d'Angleterre (Edouard de Woodstock) par le sud-ouest de la France dans le contexte de la guerre de Cent Ans et de la grande dépression médiévale.

Nous avons annoté les réponses à partir de critères simples, décrits ci-dessous :
\begin{itemize}
	\item « Bonne réponse » : contient les réponses suffisamment précises et complètes pour être totalement exploitables.
	\item « Réponse partielle » : regroupe les réponses fournissant des informations justes mais pas suffisamment précises ou complètes pour être considérées comme parfaites.
	\item « Réponse approximative » : correspond à des réponses qui ne sont pas fausses mais dont le résultat est bien trop vague et approximatif pour être exploitable.
	\item « Mauvaise réponse » : liste toutes les réponses fausses ou totalement hors-sujet.
	\item « Pas de réponse (mais indication) » : cible les absences de réponse pour lesquels les outils ont tenté de fournir une explication ou une alternative.
	\item « Aucune réponse » : regroupe les retours de LLM sans indication, les absences de réponses mais aussi les éventuels bogues.
\end{itemize}
Prenons un exemple pour clarifier cette échelle de validation. Lorsque nous posons la question contextualisée « Dans quelle ville s'est tenue la bataille de 1356 ? », nous considérons que «~Nouaillé-Maupertuis », « Nouaillé » ou « Maupertuis » sont de bonnes réponses, «~près de Poitiers » est partiellement juste, « en Nouvelle-Aquitaine » et « dans l’ouest de la France » sont trop vagues et « Châtellerault » est erroné. Cette approche permet ainsi de vérifier l'exhaustivité des réponses. En outre, selon ce que l’on souhaite obtenir \textit{a minima} comme niveau de réponses, nous pouvons regrouper des sous-types de validation. Par exemple, seules les bonnes réponses semblent parfaitement exploitables si nous souhaitons créer un outil pour des historiens, tandis que les réponses partielles voire approximatives peuvent suffire si nous attendons uniquement quelques indications sur un sujet précis. Dans nos travaux, nous avons opté pour le premier cas avec un taux de précision focalisé uniquement sur les bonnes réponses.

Outre l’évaluation de la qualité des réponses générées, nous avons pris en considération celles retournées dans une autre langue que le français. En effet, qu’une réponse soit correcte ou partiellement correcte est positif, mais si cette dernière est en anglais, elle peut devenir plus difficilement exploitable, notamment lorsque le LLM ciblé répond aléatoirement en français ou en anglais. Rappelons que nos objectifs sont à la fois de vérifier le niveau de connaissances historiques des LLM et par ailleurs de savoir s’il serait possible d’exploiter les données retournées. Bien que le fond des réponses possède une valeur majeure pour les historiens, la forme revêt également une importance capitale si nous envisageons l’usage des données obtenues. Cette analyse ne remet donc pas en cause la qualité de certains outils dans leur connaissance en histoire, mais plutôt leur fiabilité à répondre dans une langue donnée (seulement 84,33 \% des réponses ont été obtenues en français). Aussi, les LLM arrivent parfois à répondre correctement à la question mais se perdent souvent dans de longues tirades, au point d’insérer malencontreusement une réponse contradictoire voire fausse. Dans de tels cas, seule l’analyse humaine peut permettre de trancher entre les différentes catégories d’évaluation, classant ainsi ce qui aurait été une bonne réponse dans une autre catégorie, selon la gravité des anomalies générées par les modèles de langage.

\section{Résultats de l'expérience}
\label{section-resultat}
\subsection{Analyse}
L'évaluation humaine nous a permis de mesurer la justesse des réponses obtenues sur l'ensemble des questions posées, à la fois par thématique mais aussi de façon générale. Notre banc d’essai obtient ainsi une précision moyenne de seulement 33,60 \% si l’on se focalise uniquement sur les bonnes réponses, avec respectivement quatre LLM dominants dont la précision atteint au mieux 54,29 \% pour GPT-4 et au pire 45,90 \% pour Bard (Tableau~\ref{Tableau3}). Ces résultats sont plus nuancés si l’on considère que les réponses partielles ont un intérêt, avec un taux de précision global de 52,67 \% et le même quatuor dominant, cette fois-ci avec TextCortex AI en tête et ses 72,39 \% de réponses acceptables. Par ailleurs, les résultats fluctuent beaucoup pour une même question ou une nouvelle génération de réponses. En effet, nous avons testé huit requêtes par LLM (quatre déclinaisons générées deux fois) pour chaque question afin d’obtenir une certaine homogénéité dans la précision des réponses. Malheureusement, obtenir une bonne réponse n’est en rien l’assurance d’une précision équivalente d’autres variantes d’une même question, et ce phénomène se produit très régulièrement. En outre, la taille des corpus d’entraînement et le nombre de paramètres d’un LLM ne sont pas nécessairement un gage de réussite, comme le montrent les différences sensibles de précision entre les meneurs GPT-4, TextCortex AI, GPT-3.5 et Bard dont les caractéristiques sont pourtant très variables à l’origine (Tableau~\ref{Tableau2}). Par conséquent, notre analyse globale indique que même les meilleurs LLM ne garantissent pas des résultats convenables et exploitables pour des questions d’histoire.

\begin{table}[htb]
\center\small
\begin{tabular*}{\linewidth}{@{\extracolsep{\fill}}ccccc}
\hline\hline
& \textbf{Bonnes réponses} & \textbf{Autres réponses} & & \textbf{Précision} \\
\hline
ChatGPT (GPT-4) & 291 & 245 & & 54,29\% \\
ChatGPT (GPT-3.5-Turbo) & 276 & 260 & & 51,49\% \\
Google Bard & 246 & 290 & & 45,90\% \\
TextCortex AI & 277 & 259 & & 51,58\% \\
Guanaco & 183 & 353 & & 34,14\% \\
Vicuna & 197 & 339 & & 36,75\% \\
GPT4All & 95 & 441 & & 17,72\% \\
Koala & 120 & 416 & & 22,39\% \\
Vigogne & 94 & 442 & & 17,54\% \\
Falcon & 22 & 514 & & 4,10\% \\
\hline
\textbf{Totaux} & 1801 & 3559 & \textbf{Moyenne} & 33,60\% \\
\hline
\end{tabular*}
\caption{Résultats obtenus et précision des réponses générées par les LLM.}
\label{Tableau3}
\end{table}

Notre méthodologie ouvre aussi plusieurs niveaux de lecture pour ces résultats. D’abord, il est à noter que l’incidence du type de requêtes, en langue naturelle ou en mots clés, est dérisoire au niveau de la qualité des réponses. Il en va de même pour le choix des thématiques qui entraîne peu de confusions, alors que nous avions volontairement construit des questionnements ambigus autour des batailles de Poitiers, des guerres de religion ou des sièges de La Rochelle, multiples dans l’histoire. En revanche, notre banc d’essai démontre que le niveau de profondeur d’un thème historique peut nettement influencer la qualité des réponses. Par exemple, nos questions portant sur l’artisanat en Poitou durant l’époque moderne n’obtiennent que 18,17 \% de précision, tandis que nos interrogations sur des sujets moins équivoques décrochent \textit{a minima} le double de bonnes réponses (Tableau \ref{Tableau4}). Cela s’explique certainement par l’incapacité des LLM à compiler des informations complexes et à produire des réflexions autour d’un sujet nébuleux ou abstrait, alors qu’ils ont la faculté d’exploiter plus facilement des données précises à propos de thématiques mieux bornées et plus facilement intelligibles. Cette idée se confirme aussi car le type de questions posé influence assez nettement les résultats. En effet, les questions ouvertes entraînent 33,69 \% de réponses approximatives ou partiellement justes auxquelles s’ajoutent environ 25 \% de mauvaises réponses, principalement parce que les LLM ont la malencontreuse habitude de fabriquer de longues proses généralistes, sans se focaliser sur la spécificité des questions posées et sans mener une réflexion approfondie. De même, les questions quantitatives posent énormément de problèmes aux modèles de langage, avec seulement 17,42 \% de bonnes réponses trouvées, tout comme les questions fermées dont la précision est famélique (8,63 \%) lorsqu’elles attendent des listes de données en retour.

\begin{table}[htb]
\center\small
\begin{tabular*}{\linewidth}{@{\extracolsep{\fill}}cccc}
\hline\hline
\textbf{Thématiques} & \textbf{Bonnes réponses} & \textbf{Autres réponses} & \textbf{Précision} \\
\hline
Bataille de Poitiers (732) & 376 & 664 & 36,15 \% \\
Bataille de Poitiers (1356) & 425 & 615 & 40,87 \% \\
Troisième guerre de religion (1568-1570) & 395 & 645 & 37,98 \% \\
Siège de La Rochelle (1627-1628) & 387 & 653 & 37,21 \% \\
Artisanat à l'époque moderne & 218 & 982 & 18,17 \% \\
\hline
\end{tabular*}
\caption{Résultats obtenus et précision des réponses générées par thématique historique.}
\label{Tableau4}
\end{table}

Notre travail s’intéresse aussi à l'extraction des données intéressantes au sein des réponses. Bien que la précision et l'exhaustivité des réponses restent la priorité de notre étude, la composition des réponses formulées revêt une importance capitale si nous souhaitons extraire des informations historiques essentielles. Or, trois principaux problèmes se posent pour mener pleinement cette tâche sans accroc. D’une part, certains LLM comme Falcon ou GPT4All présentent quelques anomalies, nécessitant parfois plusieurs tentatives de génération avant d’obtenir une réponse. Malgré ces efforts, 1,03 \% des requêtes n'ont pas obtenu de réponses, et ce manque de fiabilité s'élève à 5,15 \% si nous ajoutons les retours uniquement indicatifs, qui ne répondent donc pas aux questions initiales. D’autre part, Koala et GPT4All répondent régulièrement en anglais, tandis que d'autres LLM restituent parfois des réponses dans une autre langue (anglais, allemand, espagnol) ou en franglais, quand bien même il leur est précisé de répondre exclusivement en français. Enfin, la tendance des modèles de langage à générer des réponses verbeuses ou à halluciner ne permet pas toujours d’extraire uniquement les données majeures et justes du propos. Pis encore, certains modèles ne comprennent parfois absolument pas une question et répondent de manière totalement délirante, comme c’est le cas pour la question quantitative « Combien de batailles ont eu lieu au cours des guerres de Rohan ? » portant sur les rébellions huguenotes (1621-1629) pour laquelle tous les LLM sauf Bard ont évoqué la « Terre du milieu » de la saga \textit{Le seigneur des Anneaux} de J.R.R. Tolkien. En définitive, si l’on cumule ces problèmes de forme aux faibles taux de précision des divers LLM testés, il nous semble impossible en l’état de pouvoir tirer profit, même partiellement, des réponses à des questions d’histoire.

\subsection{Limitations}
Nos résultats sont concluants à partir de notre batterie de test mais il demeure malgré tout quelques limites que nous devons prendre en considération :
\begin{enumerate}
    \item Excepté Bard ou ChatGPT, les modèles utilisés ne sont pas nécessairement les plus importants et puissants du marché, avec pour conséquence de potentielles différences de résultats pour certaines questions d’histoire. Plusieurs contraintes techniques et matérielles ne nous ont pas toujours permis d’utiliser des modèles dépassant les 13 milliards de paramètres\footnote{À titre d’exemple, nous n’avons pas pu utiliser le modèle Cédille entraîné spécifiquement sur du français avec 6 milliards de paramètres car il demandait trop de ressources (\http{https://arxiv.org/abs/2202.03371}).}. On peut toutefois relativiser notre propos car des modèles restreints comme TextCortex, Vicuna ou Guacano s'en sont très bien sortis dans notre étude. À l'avenir, il serait toutefois intéressant de compléter cette étude avec d'autres LLM importants comme Claude, PanGu-{\textSigma} ou Mistral-7b, ou même Ernie, YaLM, Command ou MPT si ces derniers s'adaptent à notre langue.
    \item Il eut été intéressant de tester d'autres sujets historiques et questions afin d'offrir davantage de portée et de variété à notre analyse, et de confirmer ou nuancer nos résultats.
	\item Plusieurs études démontrent qu’ajouter une chaine de pensée dans l’invite (prompt) comme « Let's think step by step » aide les modèles à fournir naturellement de meilleures réponses avec des raisonnements et étapes intermédiaires (\cite{wei_chain--thought_2023, kojima_large_2023}). Nous avons privilégié la méthode zero-shot qui consiste à poser des questions brutes comme dans n’importe quel outil de recherche. Or, peut-être que l’usage de chaînes de pensée pourrait faciliter la tâche des LLM pour certaines questions.
	\item Nos résultats ont été obtenus uniquement par une analyse manuelle et humaine tant certaines questions d'histoire font appel à des notions implicites, complexes et détaillées. L'évaluation par des experts appelle toutefois à la vigilance car les résultats peuvent être quelque peu affectés par l'introduction de biais ou de différences d'interprétation et de notation des réponses. De ce fait, si cela semble être une solution convenable pour des questions ouvertes, une méthode automatisée permettrait certainement de vérifier \textit{a minima} les réponses fermées sans s'exposer aux risques précédemment évoqués.
	\item Nous avons calculé uniquement la précision car nous estimions que l'exhaustivité se vérifiait essentiellement par les critères de notation (réponse partielle, approximative, etc.) et que la fiabilité reposait plutôt sur les possibilités d'exploitation des réponses. Cependant, une métrique de fiabilité pourrait par exemple mesurer également la capacité des LLM à produire des réponses avec le même niveau de qualité à chaque fois que l'on pose des questions similaires.
\end{enumerate}

\section{Conclusion}
\label{section-conclusion}
Notre méthode d’analyse manuelle a permis de vérifier des hypothèses concernant les capacités de dix LLM à comprendre et répondre à des questions d’histoire de types, thèmes et niveaux de difficulté différents. Nos résultats révèlent que les LLM n’offrent pas suffisamment de garantie et de constance dans les réponses produites. Certes, certains outils se montrent parfois plus efficaces sur certaines thématiques historiques mais globalement, le phénomène d’hallucination est régulièrement présent et surtout, les résultats pour des formulations différentes d’une même question sont trop changeants pour nous permettre de faire pleinement confiance aux réponses générées.

En outre, l'un des objectifs de notre analyse renvoyait à la possibilité d’exploiter les données restituées par les LLM. Les résultats produits ne sont pas toujours pleinement exploitables et ne nous permettent pas d’extraire aisément les éléments de réponse les plus pertinents. En effet, les modèles de langage manifestent une certaine incontinence verbale, à travers des phrases longues, alors que nous attendons plutôt des réponses concises et efficaces. Les données essentielles dont on souhaiterait tirer profit se retrouvent entremêlées avec des phrases à l’intérêt dérisoire. Sans une nette amélioration de la précision des réponses fournies, les LLM et leurs connaissances natives ne semblent pas permettre à ce stade un usage intensif et en toute confiance en faveur des historiens ou des humanistes numériques.

La technologie des LLM laisse toutefois entrevoir de belles promesses lorsque cette dernière est suffisamment entraînée sur un sujet. Nous pourrions donc envisager l’entraînement d’un LLM spécifique avec des sources historiques étayées et complètes. Cependant, une telle opération nécessiterait de limiter la multiplication des savoirs car notre étude a montré que l’abondance de connaissances d’un LLM entraîne généralement d’importantes confusions et incertitudes dans ses réponses. Cette piste sérieuse fournirait vraisemblablement de meilleurs résultats, sans justification et commentaire complémentaire inadaptés ou indésirables.

\bibliography{egc}

\end{document}